\documentstyle[11pt,moriond,epsfig]{article}

\def\Journal#1#2#3#4{{#1} {\bf #2}, #3 (#4)}

\def\be{\begin{equation}}
\def\ee{\end{equation}}
\def\bea{\begin{eqnarray}}
\def\eea{\end{eqnarray}}

\def\ga{\mathrel{\mathchoice {\vcenter{\offinterlineskip\halign{\hfil
$\displaystyle##$\hfil\cr>\cr\sim\cr}}}
{\vcenter{\offinterlineskip\halign{\hfil$\textstyle##$\hfil\cr
>\cr\sim\cr}}}
{\vcenter{\offinterlineskip\halign{\hfil$\scriptstyle##$\hfil\cr
>\cr\sim\cr}}}
{\vcenter{\offinterlineskip\halign{\hfil$\scriptscriptstyle##$\hfil\cr
>\cr\sim\cr}}}}}

\begin{document}
\vspace*{4cm}

\title{Galactic  dust polarized emission  at high latitudes \\
 and CMB polarization}
\author{ S. K. Sethi$^{2}$, S. Prunet$^{1}$, F.R. Bouchet$^{2}$}

\address{${}^{1}$IAS, b\^at. 121, 91405 Orsay\\ 
${}^{2}$IAP, 98bis Boulevard Arago, 75014 Paris }

\maketitle\abstracts{
We estimate the   dust polarized emission
in our galaxy at high galactic latitudes, which is the
dominant foreground for measuring CMB polarization using the high
frequency instrument (HFI) aboard Planck surveyor.  We compare it with
the level of CMB polarization and conclude that,  for angular scales
$\le 1^{\circ}$, the  scalar-induced CMB polarization and
temperature-polarization cross-correlation are  much larger than  the
foreground level at $\nu \simeq 100 \, \rm GHz$.  The
tensor-induced signals seem to be at best comparable to the foreground level.} 

\section{Introduction}

The forthcoming satellite CMB projects MAP and Planck surveyor hold
 great promise for detecting the CMB polarization. A major stumbling
 block this  detection  is the unknown level of galactic
 polarized foregrounds.
In this paper, we attempt to estimate the level of dust polarized 
emission at high galactic latitudes. 
 We model this  emission  using
 the three-dimensional HI maps of the  Leiden/Dwingeloo survey at
high galactic latitudes and  the fact that the dust emission, for a wide 
range of wavelengths, has a tight correlation with the HI emission maps
of this survey~\cite{boul}. Assuming the dust grains to 
be oblate with axis ratio $\simeq 2/3$, which recent studies
support~\cite{hil}, we 
determine the intrinsic dust polarized emissivity. The 
distribution of magnetic field 
 with respect to the dust grain distribution is quite uncertain,
 we thus  consider three extreme  cases (to be described below).

\section{Method}

From the HI maps of Leiden/Dwingloo  survey \cite{har} one can
construct a statistical model of 
the three-dimensional
 dust distribution using the relation between the optical depth for
dust emission $\tau$ at $\lambda = 250 \, \rm \mu  m$  and the HI
column density $N_{\rm \scriptscriptstyle HI}$ \cite{boul}:
\begin{equation}
{\tau \over N_{\rm \scriptscriptstyle HI}} = 10^{-25} \,  \rm  cm^2
\end{equation}
Boulanger  {\em et al. \/}\cite{boul} also  showed
that the galactic dust emission spectrum can be well fitted with a Planck
spectrum with temperature $= 17.5 \, \rm K$ with emissivity proportional
to $\nu^2$. We use this spectral dependence of dust emission
throughout this paper.  The $N_{\rm \scriptscriptstyle
HI}\hbox{--}\tau$
 correlation remains good for
 $N_{HI} \le 5 \times 10^{20} \, \rm
cm^{-2}$ which is typical for high galactic latitudes.

As our aim is to construct two-dimensional maps of polarized component of dust
emission from these three-dimensional maps of unpolarized emission, we need 
the following information: (a) the intrinsic dust polarized
emissivity,  which
depends on the type and shape of the grain, (b)
 the strength and direction of magnetic field in the diffuse cloud, and (c)
the polarization reduction  factor.

{\em Intrinsic polarized emissivity}. The galactic distribution of  dust
grains can be well understood  by the silicate/graphite model, with 
the volume fraction  of graphite between  $0.25$ and $ 0.5$ of the silicates in the 
total grain volume~\cite{lee}. Assuming spheroidal grains, 
Hildebrand and Dragovan~\cite{hil} showed that the grains are oblate with the 
ratio of axis $\simeq 2/3$. Assuming no reduction of polarization, the 
intrinsic polarized emissivity is $\simeq 30 \%$ in this case.

{\em Magnetic field}. The dust grains align themselves with the 
magnetic field. To estimate the reduction of polarization from
smearing along any line of sight, one needs
to know the direction and strength of the magnetic field. There is 
great  uncertainty in the direction of the magnetic field
relative  to dust distribution as the observational evidence show 
contradictory indications~\cite{my,go}. Therefore, we consider three extreme cases:
\begin{itemize}
\item[(1)]  The magnetic field is aligned with the major axis
of the structure. This case is relevant for dust filaments aligned
with the  field. 
\item[(2)] The magnetic field lies in the plane
perpendicular to the major axis of the structure, with its direction 
random in that plane (valid e.g. for helicoidal field around filaments)
\item[(3)] The magnetic field has the same 
direction throughout the three-dimensional map. 
\end{itemize}
  Also we assume, as recent observations and 
theoretical estimates show,  that the the dust grains are
aligned   with the magnetic field
independent of the strength of the magnetic field.

{\em Polarization reduction factor}. The reduction of intrinsic polarized
emissivity due to projection on the sky can be written as~\cite{lee}:
\begin{equation}
\Phi = R F \cos^2 \gamma,
\end{equation}
where $R$ is the Rayleigh reduction factor which gives the reduction of 
polarization due to the  inclination of grain axes about the  direction of the
magnetic field. As discussed above, we assume perfect alignment of the 
dust grains with the magnetic field and therefore take $R = 1$ throughout.
The $\cos^2 \gamma$
factor accounts for the projection of the direction of polarization on the 
plane of the sky. Using the three-dimensional maps, we calculate this factor
by first estimating, for every pixel, the direction of the structure by
finding  the direction of minimum gradient in the nearest 27 pixels in
the  four nearest  velocity templates, which are taken as slices
 in 3-dimensional space.  After finding the direction of dust structure 
with respect to the plane of the velocity template, the 
direction of magnetic field can be fixed for Case (1) and (2) of the 
magnetic field distribution.  
The $\cos^2 \gamma$ term  ($\gamma$ being the angle between the direction
of magnetic field and the plane of the velocity template) 
 can then be easily computed; and by multiplying by this
factor, the
projected distribution of polarized emission is evaluated for
 every velocity template. This procedure is used to construct the projected 
distribution of the Stoke's parameters for the first two cases of 
the magnetic field distribution. For the third case (the magnetic field 
having the  same direction everywhere), we assume, for simplicity, $\cos^2
\gamma = 0.5$.  $F$  term 
is the reduction of polarization from summing the contribution of 
different directions of polarization   along any line of sight~\cite{burn}.
This factor is estimated 
directly by vectorially adding the contribution from every velocity
template along a line of sight.

\begin{figure}
 \includegraphics[scale=0.75,angle=-90]{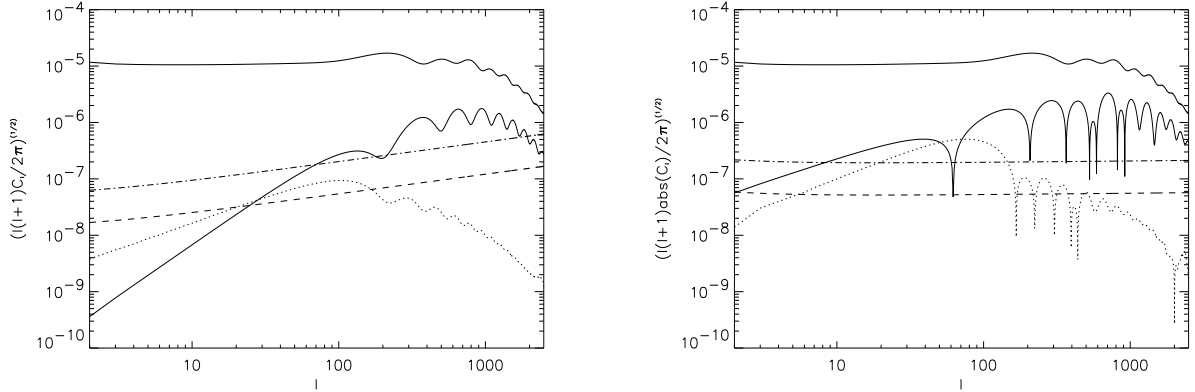}
  \caption{ Left Panel: The power spectra of
dust polarized emission is compared with the theoretical 
predictions for  CMB polarization. The scalar and tensor-induced E-mode
power spectra ({\em solid \/} and {\em dotted \/} lines respectively)
are plotted  against the dust polarized power spectra at $143 \, \rm GHz$ and 
$217  \, \rm GHz$ (the {\em dashed \/} and {\em dot-dashed \/} lines,
respectively). The CMB temperature power spectrum is shown for comparison ({\em thick
 solid\/} line). The power spectra are plotted in the units of $\Delta
T /T$. Right Panel: Same as the left panel for the $ET$ cross-correlation}.
 \label{epar}
\end{figure}

\section{Results}

\subsection{Angular distribution of dust polarized emission and
  comparison with CMB power spectra}

We construct two-dimensional  $15^{\circ}{\rm x}15^{\circ}$  maps between
latitudes $30^{\circ}$ and $75^{\circ}$ using the methods described in
the last section, and
take the Fourier transform of these  maps to obtain the power spectra of the
Stoke's parameters. 
For a comparison with  the CMB polarized component, it is convenient to 
define variables $E$ and $B$~\cite{sel}, which are linear
combinations of the usual Stoke's parameters in Fourier space. It is
because $B$ always vanishes for the
scalar-induced CMB polarization. Also the cross-correlation of $B$ with 
all the other variables vanishes.

In terms of these variables,  the  estimated power spectra and the
cross-correlation of dust polarized emission
can approximately be fitted by:
\begin{eqnarray}
C_{E}(\ell) &=& 8.9 \times 10^{-4} \ell^{-1.3} \, \,  ({\rm \mu K} )^2 \\
C_{B}(\ell) &=& 1.0 \times10^{-3}  \ell^{-1.4} \, \, ({\rm \mu K})^2 \\ 
C_{ET}(\ell) &=& 1.7 \times10^{-2}  \ell^{-1.95} \, \, ({\rm \mu K})^2. 
\end{eqnarray}
The power spectra are normalized at $100 \, \rm GHz$ and are for the  
maps between galactic latitudes $30^{\circ}$ and $45^{\circ}$ 
(where the level of contamination is the highest)  for Case
(1) of the magnetic field distribution. The other two cases of
magnetic field distribution  result in smaller  or comparable level
of foregrounds at $\ell \ge 200$ (for details, see Prunet
{\em et al \/}~\cite{prunet}).

\begin{figure}
  \resizebox{8cm}{5cm}{\includegraphics{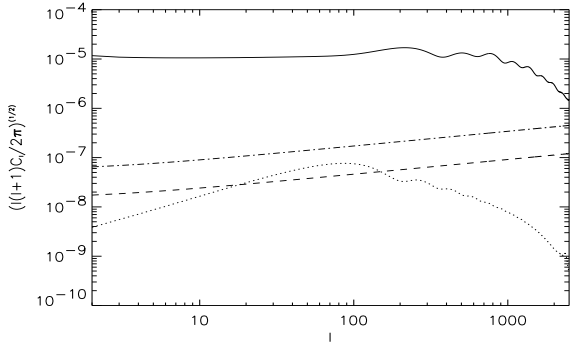}}
 \caption{Same as Fig. \ref{epar}  for B-mode power spectra.}
 \label{bpar}
\end{figure}

We  plot these against the scalar and tensor induced CMB power spectra
for various variables in Fig.~\ref{epar} and  \ref{bpar}. 
The cosmological model we adopted to compute the CMB spectra is
a flat, tilted CDM model, which generates tensor-induced B-mode
polarization, with scalar spectral index $n_s = 0.9$.
 We take the tensor
spectral index $n_t = 1-n_s =0.1$ with scalar to tensor quadrupole
ratio of $7(1-n_s)$. The CMB power spectra were computed using the CMB
Boltzmann code CMBFAST~\cite{sel}.

In Fig.~ \ref{epar}, the scalar-induced E-mode CMB power spectrum is seen to be above the
level of dust contamination for $\ell \ga 200$ for the two HFI 
frequency channels at $143 \, \rm GHz$ and $217 \, \rm GHz$. However
the tensor-induced E-mode   CMB fluctuations, which in any case 
constitute a  small part of the E-mode signal,    are likely to be swamped by
the foreground contamination. The  detection of B-mode anisotropies is
more interesting as it would   unambiguously determine  the presence of 
tensor-induced component in CMB fluctuations~\cite{se,kam,zal}.
As seen in Fig.~ \ref{bpar} 
  the B-mode power spectrum is at best comparable to
the foreground level in a small $\ell$-range. 
 Fig~\ref{epar} shows that the CMB  $ET$ cross-correlation power spectrum from
scalar perturbations is more than  an order of magnitude above the level
of foreground contamination, at least for $\ell \ga 100$ while the
tensor-induced $ET$ cross-correlation remains  below the
foreground signal.

As the  foregrounds  differ  from CMB in both $\nu$- and
$\ell$-dependence,  the level of 
 foreground contamination can be substantially reduced even in the presence of 
noise~\cite{bouchet,teg}. The extension of this technique to
include polarization and temperature-polarization cross-correlation, and
how it can be used to quantify errors on various power spectra,  is described
in a companion paper in this volume~\cite{pru}.

\end{document}